\documentclass{article}

\usepackage{arxiv}

\usepackage[utf8]{inputenc} 
\usepackage[T1]{fontenc}    
\usepackage{lmodern}        
\usepackage{hyperref}       
\usepackage{url}            
\usepackage{booktabs}       
\usepackage{amsfonts}       
\usepackage{nicefrac}       
\usepackage{microtype}      
\usepackage{graphicx}

\title{A Journey from Wild to Textbook Data to Reproducibly Refresh the Wages Data from the National Longitudinal Survey of Youth Database}

\author{
    Dewi Amaliah
   \\
    Department of Econometrics and Business Statistics \\
    Monash University \\
  Clayton, VIC 3800 \\
  \texttt{\href{mailto:dlamaleeah@gmail.com}{\nolinkurl{dlamaleeah@gmail.com}}} \\
   \And
    Dianne Cook
   \\
    Department of Econometrics and Business Statistics \\
    Monash University \\
  Clayton, VIC 3800 \\
  \texttt{\href{mailto:dicook@monash.edu}{\nolinkurl{dicook@monash.edu}}} \\
   \And
    Emi Tanaka
   \\
    Department of Econometrics and Business Statistics \\
    Monash University \\
  Clayton, VIC 3800 \\
  \texttt{\href{mailto:emi.tanaka@monash.edu}{\nolinkurl{emi.tanaka@monash.edu}}} \\
   \And
    Kate Hyde
   \\
    Department of Econometrics and Business Statistics \\
    Monash University \\
  Clayton, VIC 3800 \\
  \texttt{\href{mailto:hyde.kate.a@gmail.com}{\nolinkurl{hyde.kate.a@gmail.com}}} \\
   \And
    Nicholas Tierney
   \\
    Telethon Kids Institute \\
  Nedlands, WA 6009 \\
  \texttt{\href{mailto:nicholas.tierney@gmail.com}{\nolinkurl{nicholas.tierney@gmail.com}}} \\
  }

\providecommand{\tightlist}{%
  \setlength{\itemsep}{0pt}\setlength{\parskip}{0pt}}

\usepackage{longtable,booktabs,array}
\usepackage{calc} 
\usepackage{etoolbox}
\makeatletter
\patchcmd\longtable{\par}{\if@noskipsec\mbox{}\fi\par}{}{}
\makeatother
\IfFileExists{footnotehyper.sty}{\usepackage{footnotehyper}}{\usepackage{footnote}}
\makesavenoteenv{longtable}

\newlength{\cslhangindent}
\newlength{\csllabelwidth}
\newlength{\cslentryspacingunit} 
  {}%
  {\par}
\newenvironment{CSLReferences}[2] 
 {
  \setlength{\parindent}{0pt}
  \ifodd #1
  \let\oldpar\par
  \def\par{\hangindent=\cslhangindent\oldpar}
  \fi
  \setlength{\parskip}{#2\cslentryspacingunit}
 }%
 {}
\usepackage{calc}

\usepackage{tcolorbox}
\usepackage{fontawesome}
\usepackage{color, colortbl}
\usepackage{booktabs}
\usepackage{longtable}
\usepackage{array}
\usepackage{multirow}
\usepackage{wrapfig}
\usepackage{float}
\usepackage{colortbl}
\usepackage{pdflscape}
\usepackage{tabu}
\usepackage{threeparttable}
\usepackage{threeparttablex}
\usepackage[normalem]{ulem}
\usepackage{makecell}
\usepackage{xcolor}
\begin{document}
\maketitle

\begin{abstract}
Textbook data is essential for teaching statistics and data science methods because they are clean, allowing the instructor to focus on methodology. Ideally textbook data sets are refreshed regularly, especially when they are subsets taken from an on-going data collection. It is also important to use contemporary data for teaching, to imbue the sense that the methodology is relevant today. This paper describes the trials and tribulations of refreshing a textbook data set on wages, extracted from the National Longitudinal Survey of Youth (NLSY79) in the early 1990s. The data is useful for teaching modeling and exploratory analysis of longitudinal data. Subsets of NLSY79, including the wages data, can be found in supplementary files from numerous textbooks and research articles. The NLSY79 database has been continuously updated through to 2018, so new records are available. Here we describe our journey to refresh the wages data, and document the process so that the data can be regularly updated into the future. Our journey was difficult because the steps and decisions taken to get from the raw data to the wages textbook subset have not been clearly articulated. We have been diligent to provide a reproducible workflow for others to follow, which also hopefully inspires more attempts at refreshing data for teaching. Three new data sets and the code to produce them are provided in the open source R package called \texttt{yowie}.
\end{abstract}


\hypertarget{intro}{%
\section{Introduction}\label{intro}}

Statistics and data science education relies on cleaned and simplified data, suitably called textbook data, for clear examples about how to apply different techniques. An example of this is the wages data made public by Singer and Willett (2003) in their book, ``Applied longitudinal data analysis,'' which can be used to teach generalized linear models, and additionally hierarchical, mixed effects, and multilevel models. The data records hourly wages of a sample of high school dropouts from 1979-1994, along with the demographic variables, such as education and race, taken from the National Longitudinal Survey of Youth (NLSY79) (Bureau of Labor Statistics, U.S. Department of Labor 2021a).

The story from modeling the data (and as reported by Singer and Willett) is that wages increase with the length of time in the workforce, a higher level of education leads to higher wages, and that race makes a difference, on average. However, an exploratory analysis reveals that the individual experience varies a great deal from the overall average. Some individuals experience a decline in wages the longer they are in the workforce, and many experience volatility in their wages. It is for these reasons that the wages data was used to illustrate exploratory longitudinal data analysis in Ilk (2004), and was further developed into a case study for use in the teaching of exploratory data analysis at Iowa State University.

This disparity between the overall trend and the individual heterogeneity is what makes this data interesting. Textbook data sets have longevity if they have unresolved elements. The iris data (Anderson 1935) is a prime example. It has withstood the test of time because the three species cannot be perfectly classified, and so it continues to challenge researchers and instructors to do better in the analysis\footnote{It should be noted that the iris data is best replaced today with the penguins data (Horst, Hill, and Gorman 2020), which has similar qualities, is new, and does not suffer from a connection with eugenics (Stodel 2020)}. We argue that the wages data is in this class of textbook data, too, because it presents a challenge for longitudinal data analysis: how can we better summarize and explain the individual experience?

For statistics, and data science by association, it is increasingly important to reach the individual. One might think of this as a divergence of purpose -- statistics for public policy -- or statistics for the public. The two are not the same. As the world becomes more electronically connected, combating misinformation and mitigating conspiracy theories require statistics that address the individual. For example, with the wages data, the overall trend, across demographics, is a steady increase in wages, but the patterns among individuals is heterogeneous. Some individuals see a decline in their wages, some an increase, some have dramatic ups and downs from year to year, and the heterogeneity of the patterns are evident across demographics. Subject-specific variation explains most of the variation in wages, and even though the overall trend is statistically significant, it is weak. The message for public policy is that demographic profile is related to different wage patterns on average and some structural social change is desirable. However, discussing the overall trend with an individual is misleading -- more than likely their pattern is completely different. It could even be depressing for them to compare themselves to the overall average, especially if the individual's wages have a declining trend or high volatility. It would be more useful for an individual to know what percentage of the population have profiles like theirs. We argue that different summaries are needed to communicate with the public, and ones where they might be able to individually act on to change their own situation. We occasionally hear this voiced in the public media, also, for example, an article published in the Sydney Morning Herald argues there is no average Australian (Moncrief 2015). Thus the wages data provides educators with a challenge, from a methodological and a philosophical perspective, and wage experience is a topic of interest to many.

As a textbook data set, though, the wages data is outdated. The most recent year in the data is 1994, 9 years prior to when Singer and Willett (2003) was published. Teachers of statistics need contemporary data sets to show how techniques are relevant for today's students. Using tired old textbook data sets can imbue a misconception that the field is not current. The wages data is extracted from NSLY79, one of the best examples of open data (see details at Open Knowledge Foundation 2021), which is constantly being updated. It should be possible to continuously refresh the textbook data from the data repository. This paper describes our (non-glamorous) journey from open wild data to textbook data.

This paper demonstrates the steps of cleaning data, including subjective decisions made on dealing with anomalies, and documents the process, as recommended by Huebner, Vach, and Cessie (2016). They emphasize that making the data cleaning process accountable and transparent is imperative and essential for the integrity of downstream statistical analyses and model building. Clean data often then goes through an ``initial data analysis'' (IDA) (Chatfield 1985), where one would summarize and scrutinize the data, especially to check if the data is consistent with assumptions required for modeling. This stage is related to exploratory data analysis (EDA), coined by Tukey (1977) with a focus on learning from data. EDA can be considered to encompass IDA. In practice, the three stages of cleaning, summarizing, and exploring are cyclical, one often needs to do more cleaning after scrutinizing. Dasu and Johnson (2003) say that data cleaning and exploration is a difficult task and typically consumes a large percentage of the time spent in analyzing data.

Our approach to cleaning builds heavily on the \texttt{tidyverse} approach (Wickham et al. 2019). The data is first organized into ``tidy data'' (Wickham 2014) and then further wrangled using step-wise piping with a split-apply-combine strategy for mutating new variables (Wickham 2011). Tidy data shouldn't be confused with ``tame data,'' which Kim, Ismay, and Chunn (2018) coined to refer to textbook data sets suitable for teaching, particularly teaching statistics. The resulting (tame) data is provided in a new R package called \texttt{yowie}, which includes the code so that the process is reproducible and could be used to further refresh the data as new records are made available in the NLSY79 database.

This paper is structured in the following way. Section \ref{database} describes the NLSY79 data source. Section \ref{cleaning} presents the steps of cleaning the data, including getting and tidying the data from the NLSY79 and IDA to find and repair anomalies. Our final subset is compared to the old textbook subset in Section \ref{compare}. Finally, Section \ref{summary} summarizes the contribution and makes recommendations for the NLSY79 data curators.

\hypertarget{database}{%
\section{The NLSY79}\label{database}}

Singer and Willett (2003) used the wages and other variables of high school dropouts from the NLSY79 data as an example data set to illustrate longitudinal data modeling of wages on workforce experience, with covariates education and race. This data has been playing an important role in research in various disciplines, including but not limited to economics, sociology, education, public policy, and public health for more than a quarter of the century (Pergamit et al. 2001). In addition, this is considered a carefully designed longitudinal survey with high retention rates, making it suitable for life course research (Pergamit et al. 2001; Cooksey 2017). According to Cooksey (2017), thousands of articles and hundreds of book chapters and monographs have utilized this data. Moreover, the NLSY79 is considered the most widely used and most important cohort in the survey data (Pergamit et al. 2001).

Our aim is to refresh the wages textbook data and append it with data from 1994 through to the latest data reported in 2018, a purpose consistent with Grimshaw (2015)'s statistics education goal of embracing authentic data experiences. Here, we investigate the process of getting from the raw NLSY79 data to a textbook data set as similar as possible to that provided by Singer and Willett (2003). We should also note that race is a variable in the original data set, and for compatibility, it is also provided with the refreshed data for the \emph{purposes of studying racism, not race} (Fullilove 1998). There are a number of data sets provided by Singer and Willett (2003), but we focus only on wages data because it has captivated our attention for a number of years in our own teaching of longitudinal data analysis.

\hypertarget{database-1}{%
\subsection{Database}\label{database-1}}

The NLSY79 is a longitudinal survey administered by the U.S Bureau of Labor Statistics that follows the lives of a sample of American youth born between 1957-1964 (Bureau of Labor Statistics, U.S. Department of Labor 2021a). The cohort originally included 12,686 respondents aged 14-22 when first interviewed in 1979. For a variety of reasons, some structural, the number of respondents dropped to 9,964 after 1990. The surveys were conducted annually from 1979 to 1994 and biennially thereafter. Data are currently available from Round 1 (1979 survey year) to Round 28 (2018 survey year).

Although the main focus area of the NLSY is labor and employment, the NLSY also covers several other topics, including education, training, achievement, household, geography, dating, marriage, cohabitation, sexual activity, pregnancy, fertility, children, income, assets, health, attitudes and expectations, crime, and substance use.

There are two ways to conduct the interview of the NLSY79, which are face-to-face or telephone interviews. In recent survey years, more than 90 percent of respondents were interviewed by telephone (Cooksey 2017).

\hypertarget{target}{%
\subsection{Target data}\label{target}}

The NLSY79 data used in Singer and Willett (2003) contains the longitudinal records of male high school dropouts who first participated in the study at age 14-17 years from 1979 through to 1994. This dataset contains several variables as follows:

\begin{enumerate}
\def\labelenumi{\arabic{enumi}.}
\tightlist
\item
  ID: the respondents' ID.
\item
  EXPER: stands for experience, the temporal scale, i.e., the length of time (years) in the workforce, starting on the respondents' first day at work.
\item
  LNW: natural logarithm of wages, adjusted with 1990's inflation rate.
\item
  BLACK: binary variable, 1 indicates Black and 0 otherwise.
\item
  HISPANIC: binary variable, 1 indicates Hispanic and 0 otherwise.
\item
  HGC: the highest grade completed.
\item
  UERATE: the unemployment rate of the year of the survey. When missing, the variable is set to be 7.875 (the average rate).
\end{enumerate}

We refresh this data by re-creating the full data with records from survey years 1979 through to 2018 (the most recent year published). We also modify some variables. For example, we use a single categorical race variable instead of the two binary race variables, for reasons detailed below. We also include additional variables, some for the purpose of providing more options for data exploration in teaching examples: year of the survey, age of individual in 1979, whether the individual completed high school with a diploma or with a graduate equivalency degree (GED), the highest grade completed in the corresponding year of survey, the number of jobs the individual had in the corresponding year of survey, the total number of hours the individual usually works per week, the year when the individual started to work, and the number of years the individual worked. We do not attempt to refresh the unemployment rate variable.

We aim to create three datasets as follows:

\begin{enumerate}
\def\labelenumi{\arabic{enumi}.}
\tightlist
\item
  The wages data of the whole NLSY79 cohort, including females,
\item
  A separate table of the demographic data of the whole NLSY79 cohort, and
\item
  The subset of wages data in (1), which is the wages of high school dropouts' as a refreshed version of Singer and Willett (2003)'s data.
\end{enumerate}

\hypertarget{cleaning}{%
\section{Data cleaning}\label{cleaning}}

In the context of official statistics, M. van der Loo and de Jonge (2018) describe the ``statistical value chain,'' which includes various production stages of the data cleaning process as raw data (initial data as delivered originally), input data (data organized with correct type and identified variables), and valid data (data that has been cleaned and more accurately represents the intent of variables). What we have colorfully named wild data can be considered to be raw data, and valid data could be considered to be textbook data in the above statistical value chain. In this section, we outline the steps to download the raw data (Section \ref{getdata}) and then tidy the raw data into input data, specifically for the demographic variables (Section \ref{tidydemog}) and the employment variables (Section \ref{tidyemp}), so that the resulting input data can be used downstream for validating the data as described in Section \ref{ida} and Section \ref{censor}.

\hypertarget{getdata}{%
\subsection{Getting the data}\label{getdata}}

The NLSY79 data contains a large number of variables, but for our purposes, the scope required is limited to demographic profiles, wages data, and work experience. More specifically, we went to the NLSY79 database website at \url{https://www.nlsinfo.org/content/cohorts/nlsy79/get-data}, clicked on the direct link to NSLY79 data, and navigated as described in Figure \ref{fig:source-nav}.

\begin{figure*}[p]

\begin{tcolorbox}[title = Navigating the data source, fontupper=\linespread{.8}\selectfont]
\vspace{3mm}
\faDatabase\ NLSY79 (\url{https://www.nlsinfo.org/investigator/pages/search?s=NLSY79})\\\\
\vspace{3mm}
\faCheck\ The CASEID will be always be selected.  \\
\vspace{1mm}
\faCheck\ The 3 recommended demographic variable (sample ID, race and sex) were selected.  For the remaining variables, we went to the ``Variable Search" tab and select variables as follows
\begin{itemize}
\item[$\triangleright$] Education, Training and Achievement Scores
\begin{itemize}
\item[$\triangleright$] Education $\triangleright$ Summary measures $\triangleright$ All schools $\triangleright$ By year
\begin{itemize}
\item[$\triangleright$] Highest grade completed
\begin{itemize}
\item[\faCheck] All 80 variables in Highest grade completed were selected.
\end{itemize}
\end{itemize}
\begin{itemize}
\item[$\triangleright$] Dates of diploma or degree
\begin{itemize}
\item[\faCheck] All variables named Q3-8A were selected.
\end{itemize}
\end{itemize}
\end{itemize}
\item[$\triangleright$] Employment
\begin{itemize}
\item[$\triangleright$] Summary measures $\triangleright$ By job
\begin{itemize}
\item[$\triangleright$] Hours worked  
\begin{itemize}
\item[\faCheck] All 447 primary variables in Hours worked were selected.
\end{itemize}
\item[$\triangleright$] Hourly wages
\begin{itemize}
\item[\faCheck] All 156 variables in Hourly wages were selected.
\end{itemize}
\end{itemize}
\end{itemize}
\begin{itemize}
\item[$\triangleright$] Summary measures $\triangleright$ Since date of last interview $\triangleright$ Weeks worked
\begin{itemize}
\item[\faCheck] All 28 variables in Weeks worked were selected.
\end{itemize}
\end{itemize}
\begin{itemize}
\item[$\triangleright$] Employer Roster $\triangleright$ Job dates $\triangleright$ Original start date
\begin{itemize}
\item[\faCheck] Only selected the start date (Year) for the first job (E00101.02)
\end{itemize}
\end{itemize}
\item[$\triangleright$] Household, Geography \& Demographics
\begin{itemize}
\item[$\triangleright$] Demographics $\triangleright$ Basic demographics $\triangleright$ Date of birth
\begin{itemize}
\item[\faCheck] All 4 variables in Date of birth were selected. 
\end{itemize}
\end{itemize}
\end{itemize}
\begin{itemize}
\item[\faCloudDownload] To download all 742 variables selected, we then navigate to the tab ``Save / Download" then select the tab ``Advanced Download". We select the R Source code and comma-delimited datafile of selected variables with Reference Number as column headers. We name the filename ``NLSY79" and press the download button. There are also options to get control or dictionary files for SAS, SPSS and STATA. 
\end{itemize}
\end{tcolorbox}
\caption{Documented steps taken to select variables of interest and download the raw data.\label{fig:source-nav}}
\end{figure*}

The downloaded data set comes as a zip file, containing the following set of files:

\begin{itemize}
\tightlist
\item
  \texttt{NLSY79.csv}: comma separated value format of the response data,
\item
  \texttt{NLSY79.dat}: alternative text format of the response data,
\item
  \texttt{NLSY79.NLSY79}: tagset of variables that can be uploaded to the website to recreate the data set, and
\item
  \texttt{NLSY79.R}: R script for reading the data into R and converting the variable names and label into something more sensible.
\end{itemize}

We alter only the file path in \texttt{NLSY79.R} and run the script without any other alteration. This results in the initial processing of the raw data into two data sets, \texttt{categories\_qnames} (where the observations are stored in categorical/interval values) and \texttt{new\_data\_qnames} (the observations are stored in integer form).

To get the data into tidy form (Wickham 2014), it needs to comply with three rules: (i) each variable forms a column, (ii) each observation forms a row, and (iii) each type of observational unit forms a table. The raw data, \texttt{new\_data\_qnames}, violates these rules (i) and (ii) because information about an individual's multiple jobs over different years are in multiple columns. The raw data consequently has a large number of columns (742 to be specific). The values in the cell under the variables begin with \texttt{HRP} correspond to the hourly wage in dollars. A glimpse of this data show the format problems:

\begin{verbatim}
#> Rows: 12,686
#> Columns: 11
#> $ CASEID_1979 <int> 1, 2, 3, 4, 5, 6, 7, 8, 9, 10, 11, 12, 13, 14, 1~
#> $ HRP1_1979   <int> 328, 385, 365, NA, 310, NA, NA, NA, 214, NA, 337~
#> $ HRP2_1979   <int> NA, NA, NA, NA, 375, NA, NA, NA, NA, NA, 300, NA~
#> $ HRP3_1979   <int> NA, NA, 275, NA, NA, NA, NA, NA, NA, NA, NA, NA,~
#> $ HRP4_1979   <int> NA, NA, NA, NA, NA, 250, NA, NA, NA, NA, NA, NA,~
#> $ HRP5_1979   <int> NA, NA, NA, NA, NA, NA, NA, NA, NA, NA, NA, NA, ~
#> $ HRP1_1980   <int> NA, 457, 397, NA, 333, 275, 300, 394, 200, 318, ~
#> $ HRP2_1980   <int> NA, NA, 367, NA, NA, NA, NA, NA, NA, NA, NA, NA,~
#> $ HRP3_1980   <int> NA, NA, 380, NA, NA, NA, 290, NA, NA, NA, NA, NA~
#> $ HRP4_1980   <int> NA, NA, NA, NA, NA, NA, NA, NA, NA, NA, NA, NA, ~
#> $ HRP5_1980   <int> NA, NA, NA, NA, NA, NA, NA, NA, NA, NA, NA, NA, ~
\end{verbatim}

Thus, we re-arrange and wrangle the data into tidy data form, with columns corresponding to individual ID, year, job number, wage in dollars, and the demographic variables. This is done using the \texttt{tidyverse} suite of packages (Wickham et al. 2019): \texttt{tidyr} (Wickham 2020) to pivot the data into long-form, with \texttt{dplyr} (Wickham et al. 2020) and \texttt{stringr} (Wickham 2019) to create new variables from the downloaded data from the database, and code levels of factors by text wrangling. The long form of the data makes it possible to do these data transformations efficiently, and it is an intermediate step towards the final target data. The code for tidying the data are demonstrated at \url{https://numbats.github.io/yowie/articles/raw-to-input-data.html} but also described in the subsequent subsections.

\hypertarget{tidydemog}{%
\subsubsection{Tidying demographic variables}\label{tidydemog}}

Our final target data will include the demographic variables (variable names): sex (\texttt{sex}), race (\texttt{race}), age (\texttt{age\_1979}), highest grade completed reported in each round of the survey (\texttt{grade}), highest grade completed ever reported (\texttt{hgc}), highest grade completed in terms of years (as an integers), e.g., 9th grade = 9, 3rd-year college = 15, (\texttt{hgc\_i}), highest grade completed in 1979 (\texttt{hgc\_1979}) and whether the graduate equivalency diploma is obtained (\texttt{ged}).

For \texttt{sex} and \texttt{race}, we have simplified the original names provided in the raw data, \texttt{SAMPLE\_SEX\_1979} and \texttt{SAMPLE\_RACE\_78SCRN}, respectively. However, it is important to note that the use of ``sex'' is probably not correct. This information is provided by the individual, and although only has two categories, is more consistent with ``gender'' as defined in Heidari et al. (2016). When using the dataset in the classroom, the educators might include discourse on the use of terms ``sex'' and ``gender''. Particularly, measuring gender as a binary variable has the potential to fail to capture people who do not identify themselves as either male or female or create measurement error for people whose gender does not align with their sex classification. From a statistical perspective, this can make it difficult to adjust survey statistics to the population when gender is measured in different categorization in different data sources. Kennedy et al. (2020) provides a helpful discussion on these topics.

Similarly, there are issues for the \texttt{race} variable. The current US Federal guidelines (Office of Management and Budget 1997) state that there should be at least five categories for race, and an individual should be able to identify as more than one race category. In addition, ``Hispanic'' or ``Latino'' should be considered to be an ethnic group rather than a race. This level of detail is not available in the database. There is one variable with only three categories, and an individual can be a member of only one. Thus, the \texttt{race} variable is inadequate by today's standards, and educators should point to the current Federal guidelines when using this variable, and keep in mind that the purpose of any analysis is to study racism rather than race.

Furthermore, the object \texttt{new\_data\_qnames}, contains the variables \texttt{Q1-3\_A\textasciitilde{}Y\_1979} and \texttt{Q1-3\_A\textasciitilde{}Y\_1981}, which records two versions of the birth year of the respondent; this is also the case for the record of birth month (\texttt{Q1-3\_A\textasciitilde{}M\_1979} and \texttt{Q1-3\_A\textasciitilde{}M\_1981}). The record contains two versions of birth year and birth month as the survey recorded this in 1979 and 1981. We checked for consistency between the two versions and found no discrepancy where the responses were recorded in both 1979 and 1981. The age was then calculated using the birth year.

The next step is processing the highest grade completed. There are several ways to define this, and this should be reflected in the refreshed data to give some flexibility for downstream analysis. The first one is the highest grade ever completed is reported in the database and provided in the refreshed data as \texttt{hgc} and \texttt{hgc\_i}, encoded as a factor and integer variable, respectively. For each individual, there is only one value of \texttt{hgc} and \texttt{hgc\_i}. This variable is obtained from \texttt{new\_data\_qnames} with the name \texttt{HGC\_EVER\_XRND} and stored in year units (e.g., 10, 11, 12, 13, and so on), so the transformation is simply giving a new column name as \texttt{hgc\_i} and recoded as a factor to give \texttt{hgc} (e.g., 10th grade, 11th grade, and so on). We decide to include both the factor and integer variables to be used as examples in various classroom demonstrations to explain different data types say. Besides this, it can also serve as a demonstration of different type of data visualization ascribed by the data type and for teaching the modeling of cross sectional data.

The second definition is the highest grade completed that is reported in each round of the survey (provided in the refreshed data as \texttt{grade}). This value can change over time, but it should only increase. This has been included in the refreshed dataset to enable richer downstream analyses, for example, if one would like to explore how the temporal changes in education level affect the wages of individuals. Hence, different from \texttt{hgc}, this variable could be used to demonstrate longitudinal and time series analysis in the classroom. This is recorded in \texttt{new\_data\_qnames} as columns beginning with \texttt{Q3-4} and \texttt{HGC} with year as a suffix. In addition, it is also in the columns beginning with \texttt{HGCREV} reflecting revised data. We chose to use these revised values because there were fewer missing values indicating it had been more thoroughly checked. In the survey years when the revised value were not reported, i.e., in 2012, 2014, 2016, and 2018, we used the unrevised version accordingly.

The third definition is the highest grade completed in 1979 (provided in the refreshed data as \texttt{hgc\_1979}), corresponding to the value from the first round of the survey. This is calculated from the \texttt{grade} value in 1979. This best reflects the grade when the individual left school and is included in order to compare with the original data.

The next step is tidying to obtain \texttt{ged}. Along with \texttt{hgc}, \texttt{ged} is used to subset the data to the high school dropouts as in the original data. The graduate equivalency status is saved as a variable started with ``Q3-8A'' followed by the year of the survey. Thus, we only separate the year and the GED status. Although the GED status is asked in each round of the survey, we only retain the latest status of one's GED.

Finally, we get all of the demographic profiles of the NLSY79 cohort. We then save this data as \texttt{demog\_nlsy79}.

\hypertarget{tidyemp}{%
\subsubsection{Tidying employment variables}\label{tidyemp}}

Our target variables for the employment are to obtain respondent's mean hourly wage (\texttt{wage}), the number of jobs (\texttt{njobs}), the total hours of work per week (\texttt{hours}) for each survey year, the year when individual starting to work (\texttt{stwork}), the length of time (years) in workforce (\texttt{yr\_wforce}), and work experience measured as the number of years worked (\texttt{exp}). As the data only reports up to 5 jobs for each respondent, the maximum number of jobs is capped at 5.

From 1979 to 1987, \texttt{new\_data\_qnames} only contains one version of hours worked per week for each job (in the variables with names starting with \texttt{QES-52A}). From 1988 onward, we selected the total hours worked per week, including hours working from home (\texttt{QES-52D}). However, in 1993, this variable was missing for the first and last job, so we selected to use \texttt{QES-52A} instead. In addition, 2008 only had jobs 1-4 for the \texttt{QES-52D} variable, so we use only these.

The hourly wages are in the variables beginning with \texttt{HRP} in \texttt{new\_data\_qnames}. As a respondent may have multiple jobs, the \texttt{mean\_hourly\_wage} is computed as a weighted average of the hourly wage for each job with the number of hours worked for each job as weights (provided that the information on the number of hours is available); if the number of hours worked for any job is missing, then the \texttt{mean\_hourly\_wage} is computed as a simple average of all available hourly wages. Prior to computing the mean hourly wage, we undertook a number of steps to treat extreme observations as described below:

\begin{itemize}
\tightlist
\item
  If the hourly rate is recorded as 0, we set wage as missing, and
\item
  If the total hours of worked per week for the corresponding job is greater than 84 hours, we set the wage and hours worked as missing.
\end{itemize}

The number of jobs (\texttt{number\_of\_jobs}) for each respondent per year is computed from the number of non-missing values of hourly wage. In other words, even if the information of hours worked exists for a particular observation, we do not tally when the hourly wage is missing.

\begin{verbatim}
#> # A tibble: 10 x 6
#>       id  year mean_hourly_wage total_hours number_of_jobs is_wm
#>    <int> <dbl>            <dbl>       <int>          <dbl> <lgl>
#>  1     1  1979             3.28          38              1 FALSE
#>  2     1  1981             3.61          NA              1 FALSE
#>  3     2  1979             3.85          35              1 FALSE
#>  4     2  1980             4.57          NA              1 FALSE
#>  5     2  1981             5.14          NA              1 FALSE
#>  6     2  1982             5.71          35              1 FALSE
#>  7     2  1983             5.71          NA              1 FALSE
#>  8     2  1984             5.14          NA              1 FALSE
#>  9     2  1985             7.71          NA              1 FALSE
#> 10     2  1986             7.69          NA              1 FALSE
\end{verbatim}

For \texttt{stwork} variable, we only rename the column names from \texttt{new\_data\_qnames}. This variable is then used to calculate the next variable, \texttt{yr\_wforce} for each survey round, which is the year of survey (\texttt{year}) minus the year of individual started working (\texttt{stwork}). Finally, \texttt{exp} variable is derived from the number of weeks worked since the last interview indicated as variable started with \texttt{WKSWK} in \texttt{new\_data\_qnames}. To obtain the work experience since 1979, we calculate the cumulative value. As the measurement unit is weeks, we convert this to fractional year.

The employment and demographic variables are then joined. These data are further filtered to the cohort who participated in at least three rounds in the survey, the minimum observation recommended for longitudinal data in Singer and Willett (2003). We also opt to restrict the minimum number of observations so that the data can be used to demonstrate within-person variation when it is used to teach longitudinal data. However, it is worth noting that the original data does not restrict the number of observations for each individual, i.e., there are individuals with only one and two observations.

Finally, we save the resultant wage data on this cohort as \texttt{wages}. Note that we save \texttt{grade} variable in this dataset instead of \texttt{demog\_nlsy79} dataset because it is a longitudinal variable, while \texttt{demog\_nlsy79} is a cross-sectional dataset reflecting the state of individuals corresponding to the most recent round of the survey.

\hypertarget{calculated-variables-work-experience}{%
\subsection{Calculated variables: work experience}\label{calculated-variables-work-experience}}

Work experience is one of the most important variables in Singer and Willett (2003) as it indicates time and makes longitudinal analysis possible. It is desirable to calculate this rather than using the survey year for time because it more accurately reflects a person's time in the workforce. Thus, in the spirit of refreshing the data to the newest round of the survey, this variable needs to be calculated from other variables provided. It is not straightforward. We start with the definition of experience in Singer and Willett (2003).

Experience is years after entering the labor force. It represents the difference between the day an individual enters the labor force (\texttt{EMPLOYERS\_ALL\_STARTDATE\_ORIGINAL.01\textasciitilde{}Y\_XRND} in the database) relative to the date of the survey, which we call \texttt{yr\_wforce}. However, using this calculation produced numbers that do not quite match the original data.

Reading the section titled \href{https://www.nlsinfo.org/content/cohorts/nlsy79/topical-guide/employment/work-experience}{``Topical Guide to the Data''} in the guide suggests it should be calculated based on the variable ``number of weeks worked since the last interview.'' This would remove periods of unemployment which makes sense when measuring experience while actually working. Since it is only measured since the last interview, this needs to be cumulated for each survey year. This produced results more similar to the original data (as discussed further in Section \ref{takeaways}).

\hypertarget{ida}{%
\subsection{Initial data analysis}\label{ida}}

According to Huebner, Vach, and Cessie (2016), initial data analysis (IDA) is the step of inspecting and screening the data after collection to ensure the data is clean, valid, and ready to be deployed in the later analyses. This is supported by Chatfield (1985), who argues that the two main objectives of IDA are data description, which is to assess the structure and the quality of the data, and model formulation without any formal statistical inference.

In this paper, we conduct an IDA or a preliminary data analysis to assess the validity of the variable values in the cohort of data that the NLSY provides. The first step is validating numerical summaries of the raw data are the same as reported by NLSY79. This is followed by graphical summaries using methods available in \texttt{ggplot2} (Wickham 2016) and \texttt{brolgar} (Tierney, Cook, and Prvan 2020).

The respondents' ages ranged from 12 to 22 when first interviewed in 1979. Hence, we validate whether all respondents were in this range in the data we extracted. Additionally, the \href{https://www.nlsinfo.org/content/cohorts/nlsy79/intro-to-the-sample/nlsy79-sample-introduction}{NLSY} also provides the number of the survey cohort by their sex (6,403 males and 6,283 females) and race (7,510 Non-Black/Non-Hispanic; 3,174 Black; 2,002 Hispanic). To validate this, we used the \texttt{demog\_nlsy79}, i.e., the data with the survey years 1979 sample. Tables \ref{tab:age-table} and \ref{tab:gender-race-table} suggest the demographic data we had is consistent with the sample information in the database.

\begin{figure}

{\centering \includegraphics[width=1\linewidth]{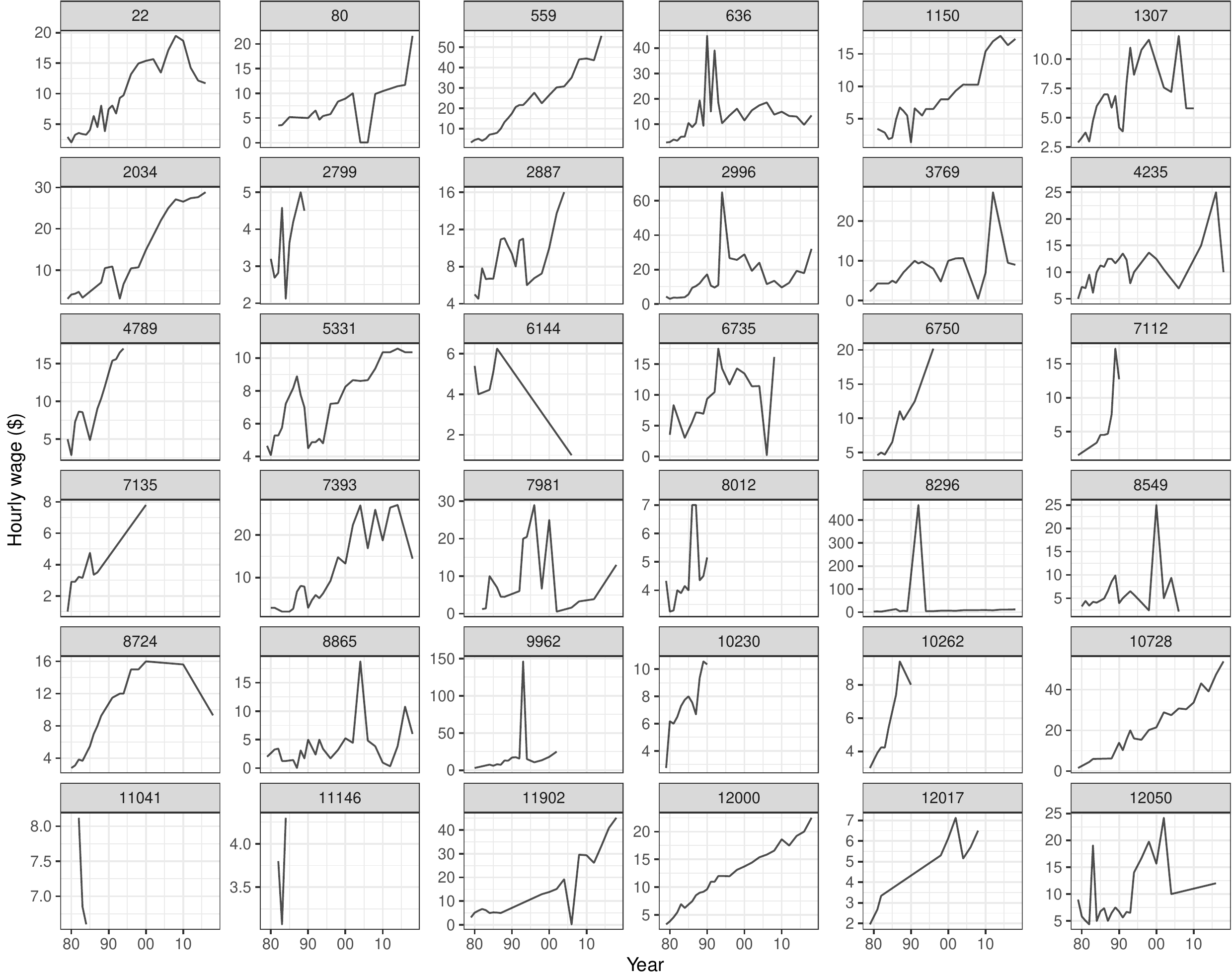} 

}

\caption{Longitudinal profiles of wages for a random sample of 36 individuals in the pre-cleaned data. There is considerable variation in wages. Some individuals  (2799, 11041, 11146) are only measured for a short period. Some individuals (8296, 9962) possibly have errors in wages in some years, because of the extreme fluctuation.}\label{fig:sample-plot}
\end{figure}

In the next step, we explore the mean hourly wage data of samples of individuals. The purpose is to examine the common patterns and check the quality. A random sample of 36 individuals is chosen (using the \texttt{sample\_n\_keys} function in \texttt{brolgar}). Their longitudinal profiles are plotted, faceted by \texttt{id}, and using free \(y\) scales so that the individual patterns can be examined (Figure \ref{fig:sample-plot}). There is a lot of variability from one individual to another and substantial fluctuation in wages at different times for most individuals. Some individuals (2799, 11041, 11146) are only measured for a short period. Some individuals (8296, 9962) possibly have errors in wages in some years because of the extreme fluctuation. These need to be inspected more closely. It is also important to note that some shorter profiles indicate some individuals have left the study before it has finished. Checking whether the demographics of the early departing are similar to those who remain in the study is an important part of any downstream analysis to account for the bias induced by the inadvertent censoring.

\begin{figure}

{\centering \includegraphics[width=1\linewidth]{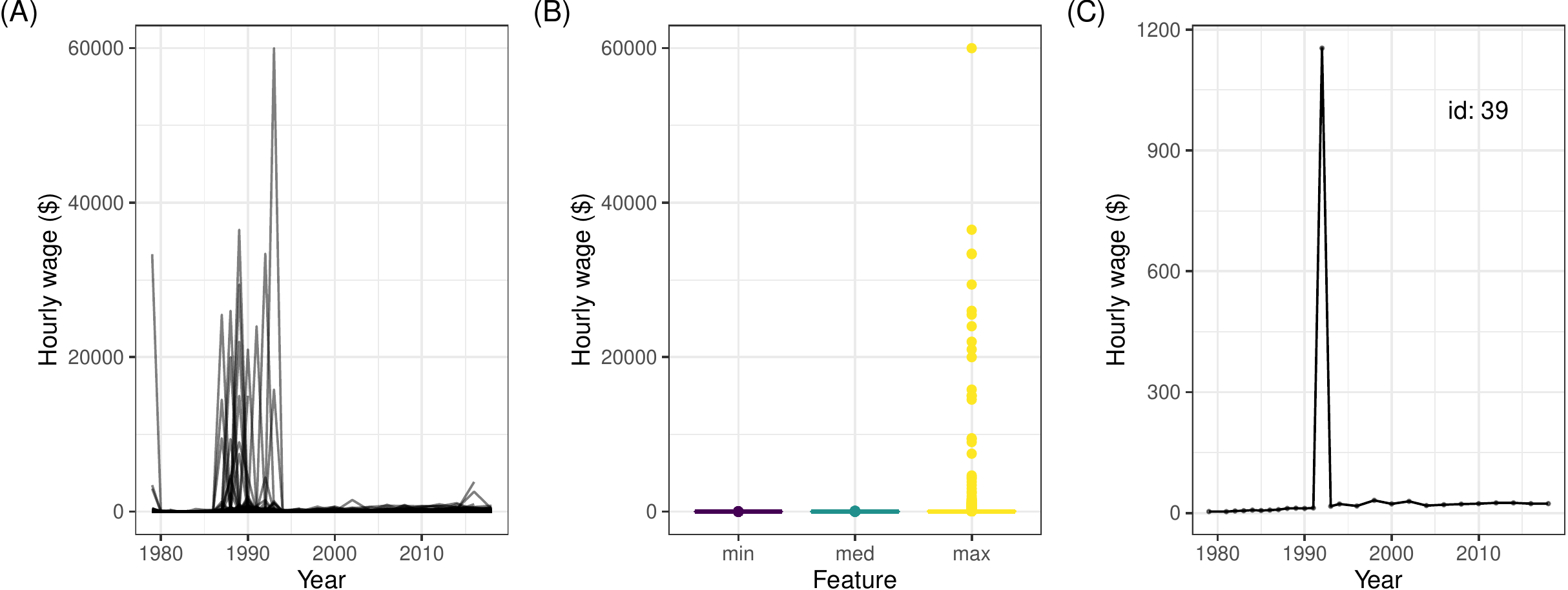} 

}

\caption{Summary plots to check the data after the tidying stage: (A) longitudinal profiles of wages for all individuals 1979-2018, (B) boxplots of minimum, median, and maximum wages of each individual, (C) and one individual (id=39) with an unusual wage relative to their years of data. It reveals that some values of hourly wages are unbelievable, and some individuals have extremely unusual wages in some years. Accordingly, more cleaning is necessary to treat these extreme values.}\label{fig:feature-plot}
\end{figure}

Figure \ref{fig:feature-plot} shows an alternative way to check the data quality. Plot (A) is the spaghetti plot where all profiles are shown, and it can be seen that there are unbelievably high wage values (up to \$60,000/hour) for some individuals, mostly around 1990. Plot (B) shows side-by-side boxplots of the three number summaries (minimum, median, and maximum) for all individuals. This tells us there are a number of individuals with unbelievably high maximum wages. Plot (C) shows the profile for an individual with a maximum wage that is not so extreme but still indicates a problem: their wages are consistently low except for one year where they earned close to \$1200/hour. This does not seem to be reasonable and leads us to use a procedure to detect and fix these temporal anomalies.

\begin{table}

\caption{\label{tab:age-table}Frequency table of the age at the start of the survey in NSLY79 cohort in the extracted data.}
\centering
\begin{tabular}[t]{rr}
\toprule
Age & Number of individuals\\
\midrule
15 & 1,265\\
16 & 1,550\\
17 & 1,600\\
18 & 1,530\\
19 & 1,662\\
20 & 1,722\\
21 & 1,677\\
22 & 1,680\\
\bottomrule
\end{tabular}
\end{table}

\begin{table}

\caption{\label{tab:gender-race-table}Contingency table for sex and race for the extracted NLSY79 demographic data. The percentage (rounded to closest 1\%) is out of the total corresponding to row.}
\centering
\begin{tabular}[t]{lrrrr}
\toprule
\multicolumn{1}{c}{ } & \multicolumn{3}{c}{Race} & \multicolumn{1}{c}{ } \\
\cmidrule(l{3pt}r{3pt}){2-4}
Sex & Hispanic & Black & Non-Black, Non-Hispanic & Total\\
\midrule
F & 1,561 (25\%) & 1,002 (16\%) & 3,720 (59\%) & 6,283\\
M & 1,613 (25\%) & 1,000 (16\%) & 3,790 (59\%) & 6,403\\
\midrule
Total & 3,174 (25\%) & 2,002 (16\%) & 7,510 (59\%) & 12,686\\
\bottomrule
\end{tabular}
\end{table}

Extremely high values were also found in the total hours of work, where some observations reported having worked for 420 hours a week in total. According to Pergamit et al. (2001), one of the flaws of the NLSY79 employment data is the NLSY79 collects the information of the working hours since the last interview. Thus, it might be challenging for the respondents to track the within-job hours' changes between survey years, especially for the respondents with fluctuating working hours or seasonal jobs. It even has been more challenging since 1994, after which respondents were only surveyed every other year and thus had to recall two full years' job history. This shortcoming might also contribute to the fluctuation of one's wages data.

\hypertarget{censor}{%
\subsection{Replacing extreme values}\label{censor}}

A robust linear regression model using the \texttt{rlm} function from \texttt{MASS} package (Venables and Ripley 2002) is used to treat the extreme values in the data. The robustness weight is used to determine if a value should be replaced with the fitted value from the model. This is constructed for each ID utilizing the \texttt{nest} and \texttt{map} function from \texttt{tidyr} (Wickham 2020) and \texttt{purrr} (Henry and Wickham 2020), respectively. An alternative approach would be a robust linear mixed model using \texttt{robustlmm} (Koller 2016).

\begin{figure}

{\centering \includegraphics[width=0.9\linewidth]{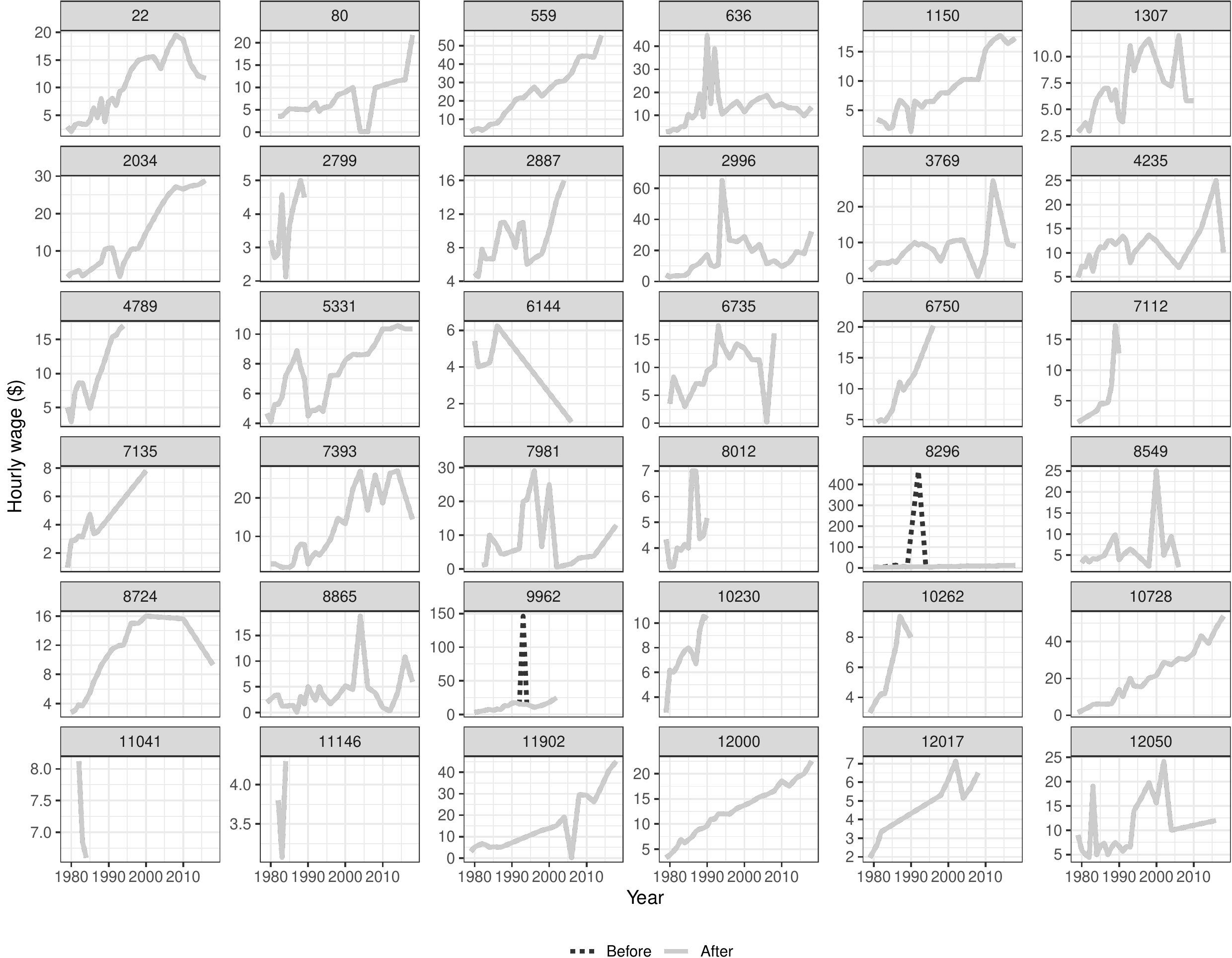} 

}

\caption{Comparison between the original (black dots) and the corrected (solid grey) mean hourly wage for same sample of individuals as shown in Figure 2. A robust linear model prediction was used to identify and correct mean hourly wages value. The extreme spikes, corresponding to implausible wages, have been replaced with values more similar to wages in neighboring years for individuals 8296 and 9962, but otherwise the profiles have not changed.}\label{fig:compare-plot}
\end{figure}

Figure \ref{fig:compare-plot} shows the profiles for a sample of 30 individuals the mean hourly wage before and after extreme values are replaced. The plot shows fluctuations in wages remain, but the large spikes (in this sample, individuals 8296, 9962), which are considered implausible, are replaced.

The challenging part of detecting an anomaly using the robustness weight is determining the weight threshold beyond which the observations are considered outliers. It is also important not to be too aggressive in outlier removal, because this would overly smooth the data, and ignore actual workforce experiences important wage volatility such as, abnormally high wages for a fleeting amount of time. To explore the risk of being overly vigorous in labeling observations as outliers, a range of threshold value were assessed. Numerous samples of individuals were drawn for each threshold, and the smoothness of the profiles was examined. Overly smooth profiles would indicate the replacement of values was too severe, resulting in removing the very interesting volatility of wages seen in many individuals. This was achieved using a purpose written \texttt{shiny} (Chang et al. 2020) app (provided with the code of this paper). Observations where a change was made are tracked with a new variable called \texttt{is\_pred}, so that this effect in the downstream analyses can be monitored.

\begin{figure}

{\centering \includegraphics[width=1\linewidth]{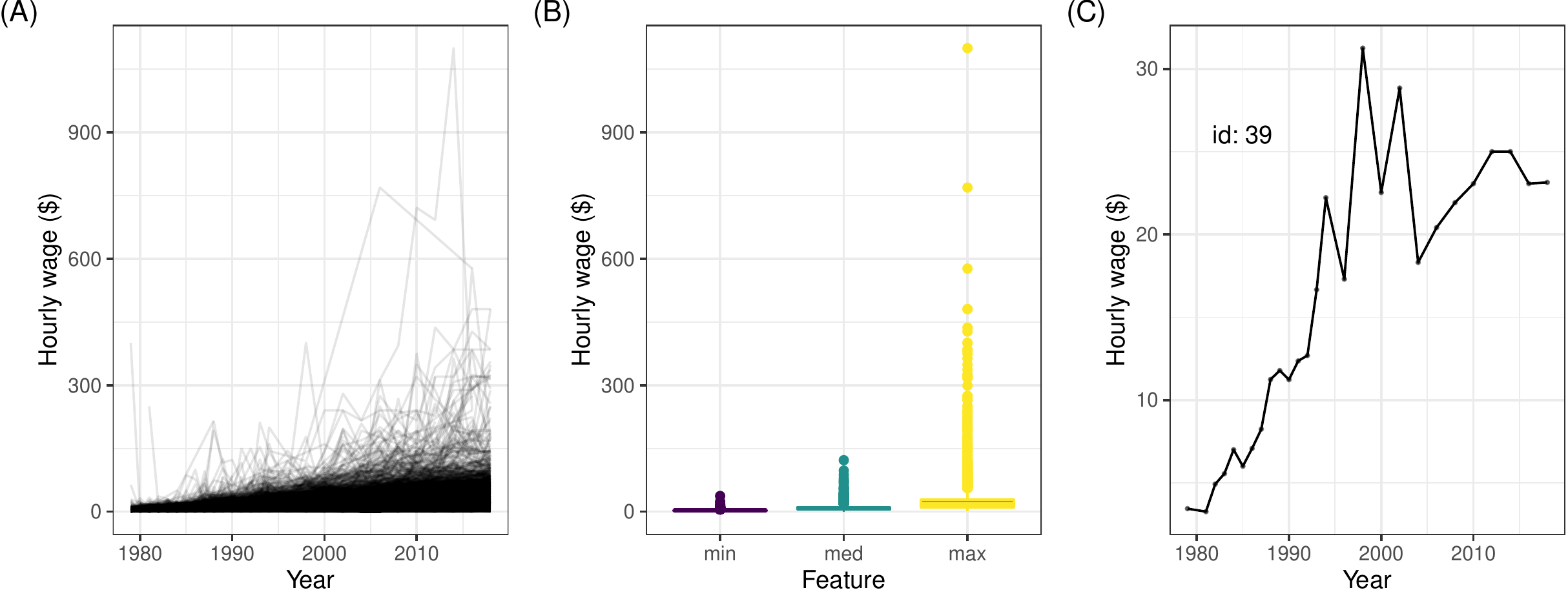} 

}

\caption{Re-make of the summary plots of the fully processed data suggest it is now in a reasonable state: (A) longitudinal profiles of wages for all individuals 1979-2018, (B) boxplots of minimum, median, (C) and maximum wages of each individual, and one individual with an unusual wage relative to their years of data. }\label{fig:fixed-feature-plot}
\end{figure}

Figure \ref{fig:fixed-feature-plot} shows the summary statistics after removing extremes. The highest wage overall is now around \$1000. Plot (A) shows a more reasonable spaghetti plot, where there are some profiles with high wages, but most profiles have wages under \$300, and there has been a steady increase in wages over the years. Plot (B) shows there are still a small number of individuals with high maximum wages. Plot (C) shows the profile for ID=39 after imputing the extreme value. The wages for this individual increase over the years, and do fluctuate some between 1900 and 2005.

\hypertarget{recap}{%
\subsection{Recap}\label{recap}}

\begin{figure}

{\centering \includegraphics[width=0.85\linewidth]{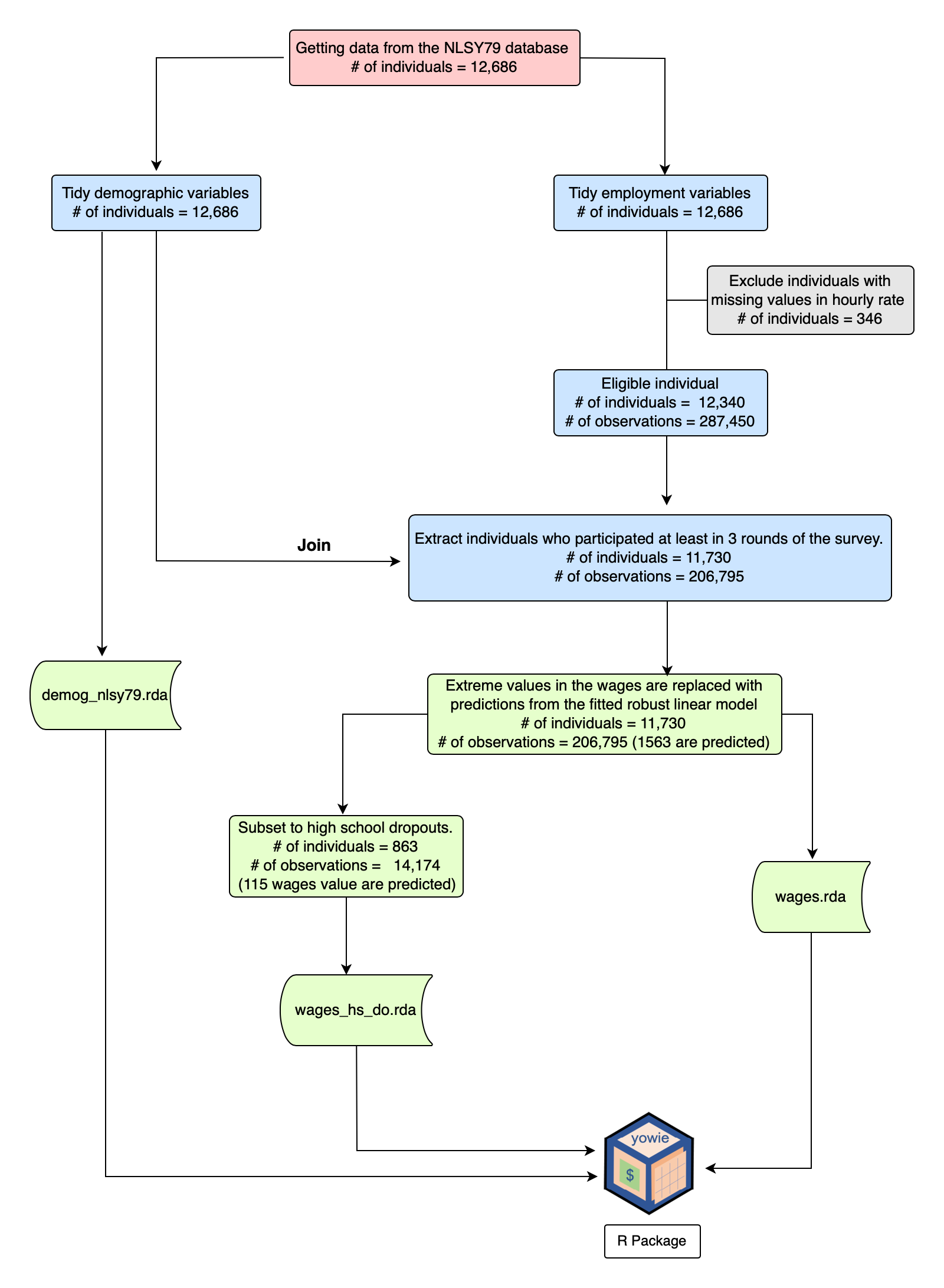} 

}

\caption{The stages of data cleaning from the raw data to get three datasets contained in \texttt{yowie}. ``\# of individuals'' means the number of respondents included in each stage, while ``\# of observations'' means the number of rows in the data. The color represents the stage of data cleaning in the statistical value chain (M. P. J. van der Loo and de Jonge 2021). Pink, blue, and green represent the raw, input, and valid data, respectively.}\label{fig:flow-chart}
\end{figure}

There are many steps and decisions made to go from raw to input to valid data. Figure \ref{fig:flow-chart} summarizes these in order to create a refreshed wages data set.

The list of variables provided in the three new datasets are as follows:

\textbf{\texttt{demog\_nlsy79}} :

\begin{enumerate}
\def\labelenumi{\arabic{enumi}.}
\tightlist
\item
  \texttt{id}: A unique individual's ID number.
\item
  \texttt{age\_1979}: The age of the individual in 1979.
\item
  \texttt{sex}: Sex of the individual, f = Female and m = Male.
\item
  \texttt{race} : Race of the individual belong to, NBH = Non-Black, Non-Hispanic; H = Hispanic; B = Black.
\item
  \texttt{hgc}: The highest grade completed ever.
\item
  \texttt{hgc\_i}: Integer value of the highest grade completed ever.
\item
  \texttt{hgc\_1979}: The highest grade completed in 1979 (integer value).
\item
  \texttt{ged}: Whether the individual had a high school diploma or Graduate Equivalency Degree (GED). 1: High school diploma; 2: GED; 3: Both.
\end{enumerate}

\textbf{\texttt{wages}} and its subset for high school dropouts cohort \textbf{\texttt{wages\_hs\_do}}:

\begin{enumerate}
\def\labelenumi{\arabic{enumi}.}
\item
  \texttt{id}: A unique individual's ID number. This is the \texttt{key} of the data as we saved the data as a \texttt{tsibble} object.
\item
  \texttt{year}: The year the observation was taken. This is the \texttt{index} of the data.
\item
  \texttt{wage}: The mean of the hourly wages the individual gets at each of their different jobs. The value could be a weighted or an arithmetic mean. The weighted mean is used when the information of hours of work as the weight is available. The mean hourly wage could also be a predicted value if the original value is considered influential by the robust linear regression as part of data cleaning.
\item
  \texttt{age\_1979}: The age of the individual in 1979.
\item
  \texttt{sex}: Sex of the individual, f = Female and m = Male.
\item
  \texttt{race}: Race of the individual belong to, NBH = Non-Black, Non-Hispanic; H = Hispanic; B = Black.
\item
  \texttt{grade}: Integer value of the highest grade completed corresponding to the survey year.
\item
  \texttt{hgc}: The highest grade completed ever.
\item
  \texttt{hgc\_i}: Integer value of the highest grade completed ever.
\item
  \texttt{hgc\_1979}: The highest grade completed in 1979 (integer value).
\item
  \texttt{ged}: Whether the individual had a high school diploma or Graduate Equivalency Degree (GED). 1: High school diploma; 2: GED; 3: Both.
\item
  \texttt{njobs}: Number of jobs that an individual has.
\item
  \texttt{hours}: The total number of hours the individual usually works per week.
\item
  \texttt{stwork}: The year when the individual started to work.
\item
  \texttt{yr\_wforce}: The length of time in the workforce in years (\texttt{year} - \texttt{stwork}).
\item
  \texttt{exp}: Work experience, i.e., the number of years of working.
\item
  \texttt{is\_wm}: Whether the mean hourly wage is weighted mean, using the hour work as the weight, or regular/arithmetic mean. TRUE = is weighted mean. FALSE = is regular mean.
\item
  \texttt{is\_pred}: Whether the mean hourly wage is a predicted value of RLM or not.
\end{enumerate}

\hypertarget{compare}{%
\section{Comparison of refreshed with the original data}\label{compare}}

The original set, containing wages of high school dropouts (Singer and Willett 2003) from 1979 through to 1994, is available in the R package \texttt{brolgar} (Tierney, Cook, and Prvan 2020). To compare the refreshed data with the original, a subset needs to be matched. There are numerous ways to do this, with the simplest being to extract the individuals based on their id being part of the original data and restricting the longitudinal measurements to the same years. However, we decided to try to replicate the process, as suggested by the description of the original data. This requires first identifying individuals who dropped out of high school.

\hypertarget{filtering-determining-who-is-a-dropout}{%
\subsection{Filtering: Determining who is a dropout}\label{filtering-determining-who-is-a-dropout}}

There is no explicit explanation of how the dropouts cohort is determined in the original data. Hence, we use the high school dropouts criteria from Wolpin (2005), which are:

\begin{enumerate}
\def\labelenumi{\arabic{enumi}.}
\tightlist
\item
  An individual whose highest grade completed (\texttt{hgc}) is reported to be less than 12th grade, \textbf{or}
\item
  An individual whose highest grade completed (\texttt{hgc}) is reported to be at least 12th grade and has received a GED (i.e.~\texttt{ged} code is 2).
\end{enumerate}

An additional criterion from Singer and Willett (2003) is to only include males aged between 14 and 17 years old in 1979. With this filtering, we obtained 670 individuals in the refreshed data compared to 888 individuals in the original data. To investigate the reason for the difference, individuals from the original and refreshed dataset were matched by id. This revealed several reasons for the disparity:

\begin{enumerate}
\def\labelenumi{\arabic{enumi}.}
\tightlist
\item
  173 individuals were more than 17 years old in 1979. Thus, it looks like the description of the original data is not quite accurate that there are people older than 17 in the subset. Our decision is to also include them in the refreshed data as the new data contains an age variable, so analysts could filter them later.
\item
  79 individuals were less than or equal to 17 years old in 1979. However, they were not captured in the refreshed data because:

  \begin{enumerate}
  \def\labelenumii{\roman{enumii}.}
  \tightlist
  \item
    35 of them completed at least 12th grade with a diploma instead of GED (\texttt{ged} variable is coded as 1). This suggests they are not dropouts, and so we excluded them from the refreshed data.
  \item
    The information about \texttt{ged} is missing in 38 individuals. We decided to include them in the refreshed data.
  \item
    3 individuals have both diploma and GED (\texttt{ged} is coded as 3). These were kept in the refreshed data.
  \item
    12 individuals do not exist in \texttt{wages} data because they have participated in less than 3 rounds of the survey.
  \end{enumerate}
\end{enumerate}

The filtering was re-applied using these decisions, resulting in a refreshed dropout subset containing 863 individuals.

\hypertarget{summaries-of-original-with-refreshed-dropouts-data}{%
\subsection{Summaries of original with refreshed dropouts data}\label{summaries-of-original-with-refreshed-dropouts-data}}

Because the original data does not have the year of collection, it is not possible to merge the two subsets directly. Merging longitudinal data requires both the key (\texttt{id}) and the index (ideally survey year). In the original data, the experience variable is the time index, and it was not possible to exactly match this for the refreshed data. Thus, comparisons of the two sets have to be conducted in a two-sample fashion rather than a matched sample.

\begin{figure}

{\centering \includegraphics[width=1\linewidth]{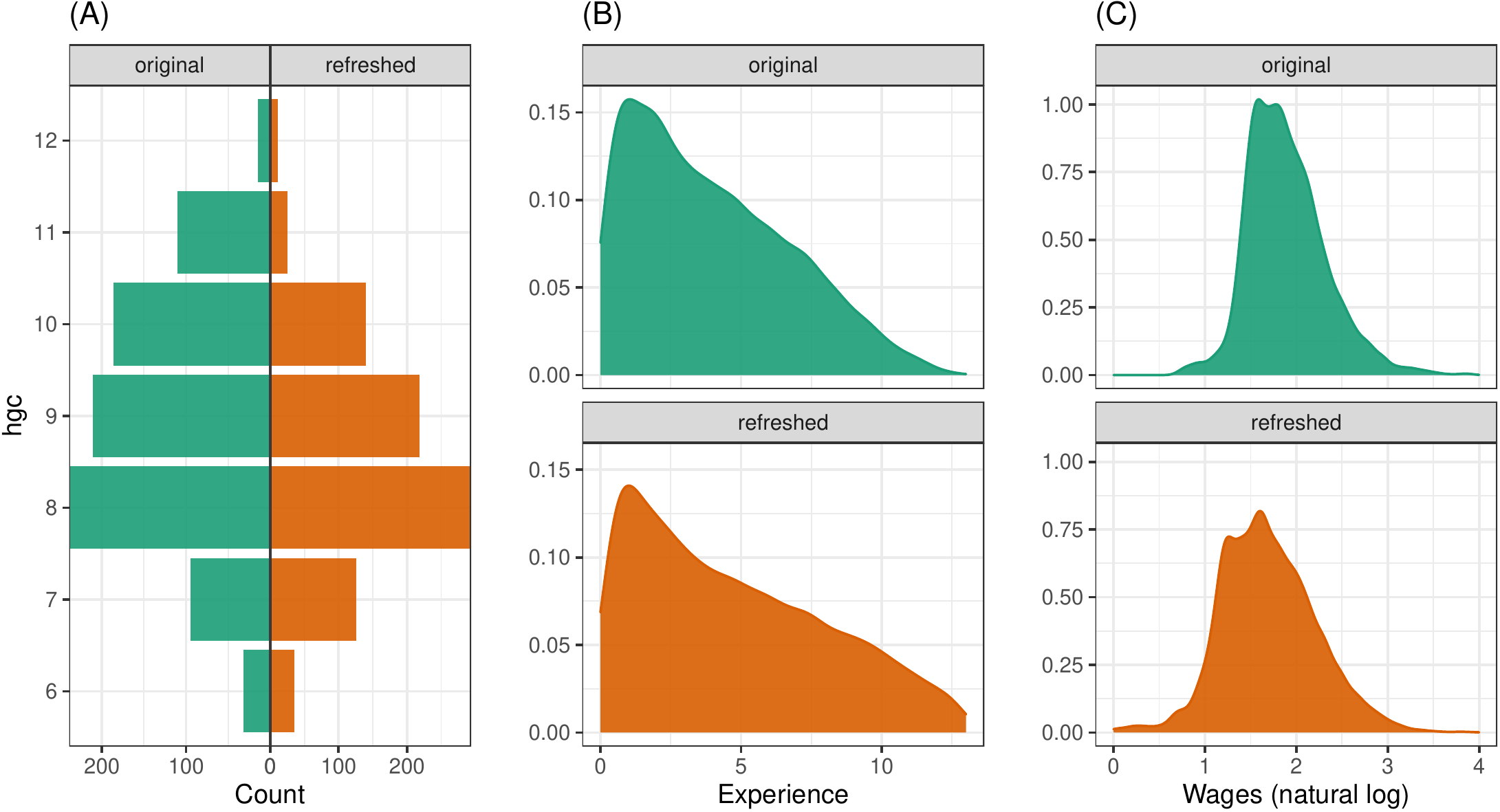} 

}

\caption{Comparison of original and refreshed data: (A) highest grade completed, (B) experience and (C) log wages. Some difference in wages would be expected because the refreshed data is not inflation-adjusted, but the two sets are reasonably similar.}\label{fig:compare-subsets}
\end{figure}

Figure \ref{fig:compare-subsets} contains summaries of corresponding variables in the two subsets. Plot (A) shows a back-to-back bar chart of the highest grade completed. The two sets are almost the same, but small differences remain. Plots (B) and (C) show stacked density plots of experience and log wages, respectively. The distributions are relatively close, with more differences in wages as would be expected because the refreshed data is not inflation-adjusted.

\hypertarget{takeaways}{%
\subsection{The takeaways}\label{takeaways}}

There are several aspects of the original data that were difficult to replicate. The calculation of the work experience is not clearly articulated. Singer and Willett (2003) describe the temporal variable as the years of experience since the first day of work. This variable is not explicitly available in the database. From the NLSY79 topical guide (Bureau of Labor Statistics, U.S. Department of Labor 2021b), we find several variables are tagged as work experience-related variables. One of them is the weeks worked since the last interview. This is used to calculate the variable. It produces reasonably similar but not exactly the same values. Because the original data set did not include the year of the survey, it cannot be precisely compared.

The highest grade completed has some confusion. There are several ways this is reported, including the highest grade ever completed and also the \texttt{hgc} at each survey year. To match the original data, it is appropriate to use the \texttt{hgc} while in high school. The documentation suggests this variable is available, but it is not actually present in the data. Hence, to match the original data, we have calculated the \texttt{hgc} to match the \texttt{hgc} achieved in the years between 1979 and 1994 based on the yearly survey value. The result does not exactly match the original for a few individuals.

In the original dataset, wages were inflation-adjusted to 1990 prices. This is not done for the refreshed data because we plan to keep refreshing it as the data is added to and released from the survey. Instead, we have provided a function in the R package, \texttt{yowie}, for users to conduct the inflation adjustment when they are ready to analyze the data.

It is important to note that the treatment of unlikely wages differs in the refreshed data. In the original data by Singer and Willett (2003), wages greater than \$75 are set to be missing. However, this value is too low to be set as the maximum threshold, and it doesn't take into account temporal neighbors for an individual. We opted to use the weights from a robust linear regression to determine what should be regarded as extreme and imputed them with their predicted values as described in Section \ref{censor}.

Lastly, the matching of individuals in the dropouts subset of the refreshed data with the original data was done using their \texttt{id}s. (Only males were in the original data.) It was comforting to see that all of the individuals in both sets do match on sex (all are male!), and race.

\hypertarget{summary}{%
\section{Summary}\label{summary}}

Any longitudinal dataset used for education should have a sufficiently reproducible process to be refreshed with new data. The NLSY79 dataset is a great teaching tool that has become outdated, both because the dataset stops in 1994, and because the demographic data could be handled more delicately. This paper has illustrated the necessary steps and decisions made to take a particular open data set and make it a textbook data set, ready for the classroom or research. In the first stage, we showed the steps performed to get the data from the NLSY79 database. The data format was converted to a tidy format for more flexibility in cleaning and exploring. Initial data analysis was conducted to investigate and screen the quality of the data. We found and provided a fix for many anomalous observations in wages using a robust linear regression model. The refreshed data is compared with the original set using a variety of numerical summaries and graphics. The current subset is made available in a new R package, called \texttt{yowie}.

The data cleaning process is documented, and the code has been made available. These provide the opportunity to again refresh the textbook data as new data is published into the NLSY79 database. Determining an appropriate robustness weight from which to threshold unusual observations was conducted using a \texttt{shiny} (Chang et al. 2020) app, which is provided with the code, and the choices used in the refreshed data are documented.

Various difficulties were encountered in trying to refresh the data, which include:

\begin{itemize}
\tightlist
\item
  Determining which records should be downloaded from the database.
\item
  Calculating experience in the workforce requires comparing the date of the first job with the first year the individual was recorded, both of which are available in the database.
\item
  Treating the extreme values since there are many unusually high hourly wages, e.g., greater than \$60,000 per hour.
\item
  Determining the dropouts subset as there is no explicit variable in the database recording high school dropout, which means we needed to compare the date of 12th grade with their GED status.
\item
  Matching IDs from the original data with those in the refreshed data do refer to the same person based on the demographic information available.
\end{itemize}

Ultimately, the refreshed data is reasonably similar to the original but unsatisfactorily far from it. The last step required would be to inflation-adjust wages. This is better to do with each wave of new data added so that it is relative to the last date in the data. Our decision was to provide the raw wages and include code to make the adjustment as part of the package.

Some readers may disagree with our decisions made to produce the refreshed textbook data and may have better insight than us in producing more appropriate textbook data. We do not assert that we have produced the best textbook data, but rather we describe our journey to provide a reasonable textbook data set. All code and documentation are provided for transparency. Readers could use this to make different decisions or provide suggestions through the package for better choices. Future updates of the \texttt{yowie} package may contain additional variables, or filters of the full set if it is deemed important.

For the data providers, we recommend a better validation system with clear rules applied at data entry and alternative output formats, such as a tidy format, which would help users better use their resources. The problem with many wages records is that there are implausible values or confusion on how to record wages for multiple jobs. These values can be validated with simple checks at data entry. Providing an open data resource is also accompanied by the responsibility that the data, especially as valuable as this, is reliable. Users need to be able to trust the data.

Why is the work presented here useful for teachers and students? There are several reasons:

\begin{itemize}
\tightlist
\item
  It primarily illustrates the steps in cleaning and processing a well-known longitudinal data so that it can be further refreshed as a textbook data set as new data emerges. Choices made during the cleaning and processing can affect findings made with the data, are transparent, and can be a basis for classroom discussions.
\item
  Some of the methods shown, and lessons learned about the data, can also be considered generally for data cleaning, initial data analysis and exploratory data analysis, useful for working with other data sets.
\item
  The refreshed data can be used for teaching advanced linear modeling, using almost identical code from Singer and Willett (2003), and thus make those lessons more current for today's students.
\item
  It provides the data to develop a case study for teaching exploratory data analysis of longitudinal data, focusing on the individual experience. The vignettes of the \texttt{brolgar} (Tierney, Cook, and Prvan 2020) package provide useful guides for the case study content.
\end{itemize}

The wages data provides a good opportunity to discuss the difference between statistics for public policy and statistics that relate to the individual. Public policy is based on models, yielding averages that might vary across strata. Modeling the wages relative to workforce experience with demographic covariates, we learn that there are significantly different patterns. More education leads to increasingly higher wages, which is a satisfying result for educators. It means education makes a difference in the wage experience, so public policy encouraging education is a data-supported action. We would also learn, although not presented here, that wages on average differ by race category. This is a disturbing finding because there is no rationale for such a difference in a fair society. This would provide support for action in public policy to remove this overall average effect. It also provides an example for educators to explain the use of data to support public policy action.

On an individual level, people might want to see how their characteristics relate to those in the dataset. To do this, the individual profiles need to be explored. Predominantly, we would learn the variation from one individual to another is far more than the variation between demographic strata. For example, many individuals with lower educational attainment earn very high wages. From a statistics and data science educator perspective, more focus and more methodology for this type of statistics need to be included in the curriculum.

The above interpretations of results from analyzing the wages data rely on trusting that the data provided is accurate and valid. The wages data is collected by a reputable organization, but we found the data has obvious errors that should be corrected. This paper has illustrated procedures and guidelines to achieve valid data and provides the code and details for it to be reproduced and modified if deemed appropriate.

Having trustworthy data is imperative for statistics and data science education. Whenever one uses a textbook data set that is perceived as relating to the students' lives, there will be interpretations made. The wages data is an example of this. Students can take away multiple interpretations from the data (e.g.~wages increase with experience, education, and race) based on the teaching focus to learn about the world. If one uses data examples that are synthetic, instilled with our own inherent prejudices (e.g., sex and race), or data that has been poorly processed containing errors, we, as educators, are being irresponsible because the societal message taught to students may be flawed. This paper demonstrates the process of producing a trustworthy data set for teaching.

\hypertarget{acknowledgements}{%
\section{Acknowledgements}\label{acknowledgements}}

We would like to thank Aarathy Babu for the insight and discussion during the writing of this paper.

The entire analysis is conducted using \texttt{R} (R Core Team 2020) in RStudio IDE using these packages: \texttt{tidyverse} (Wickham et al. 2019), \texttt{ggplot2} (Wickham 2016), \texttt{dplyr} (Wickham et al. 2020), \texttt{readr} (Wickham and Hester 2020), \texttt{tidyr} (Wickham 2020), \texttt{stringr} (Wickham 2019), \texttt{purrr} (Henry and Wickham 2020), \texttt{brolgar} (Tierney, Cook, and Prvan 2020), \texttt{patchwork} (Pedersen 2020), \texttt{kableExtra} (Zhu 2019), \texttt{MASS} (Venables and Ripley 2002), \texttt{janitor} (Firke 2020), and \texttt{tsibble} (Wang, Cook, and Hyndman 2020). The paper was generated using \texttt{knitr} (Xie 2014) and \texttt{rmarkdown} (Xie, Dervieux, and Riederer 2020).

\hypertarget{supplementary-materials}{%
\section{Supplementary Materials}\label{supplementary-materials}}

\begin{itemize}
\item
  \textbf{Code}: R script to reproduce data tidying and cleaning is available at \url{https://numbats.github.io/yowie/articles/process-data.html}. The code for the extreme value handling with robust linear fits is in \url{https://numbats.github.io/yowie/articles/input-to-valid-data.html}.
\item
  \textbf{R Package}: \texttt{yowie} is an R data package that contains 3 datasets, namely the high school mean hourly wage data, high school dropouts mean hourly wage data, and demographic data of the NLSY79 cohort. This package can be accessed from \url{https://github.com/numbats/yowie}.
\item
  \textbf{shiny app}: An interactive \texttt{shiny} web app to visualise the effect of selecting different weight threshold for substituting the wages data to its predicted value from a fit of the robust linear regression model. This app can be accessed at \url{https://ebsmonash.shinyapps.io/yowie_app/} with the source code provided \url{https://github.com/numbats/yowie/tree/master/app}.
\end{itemize}

\hypertarget{data-availability-statement}{%
\section{Data Availability Statement}\label{data-availability-statement}}

The authors confirm that the data supporting the findings of this study are available within the supplementary materials.

\hypertarget{references}{%
\section*{References}\label{references}}
\addcontentsline{toc}{section}{References}

\hypertarget{refs}{}
\begin{CSLReferences}{1}{0}
\leavevmode\vadjust pre{\hypertarget{ref-iris-data}{}}%
Anderson, Edgar. 1935. {``The Irises of the Gaspe Peninsula.''} \emph{Bulletin of the American Iris Society} 59: 2--5.

\leavevmode\vadjust pre{\hypertarget{ref-nlsy79}{}}%
Bureau of Labor Statistics, U.S. Department of Labor. 2021a. {``National Longitudinal Survey of Youth 1979 Cohort, 1979-2016 (Rounds 1-28).''} Produced and distributed by the Center for Human Resource Research (CHRR), The Ohio State University. Columbus, OH, through \url{https://www.nlsinfo.org/bibliography-citing-nls-data}.

\leavevmode\vadjust pre{\hypertarget{ref-nlsy79guide}{}}%
---------. 2021b. {``National Longitudinal Survey of Youth 1979 Cohort, Topical Guide to the Data.''} \url{https://www.nlsinfo.org/content/cohorts/nlsy79/topical-guide/employment/work-experience}.

\leavevmode\vadjust pre{\hypertarget{ref-shiny}{}}%
Chang, Winston, Joe Cheng, JJ Allaire, Yihui Xie, and Jonathan McPherson. 2020. \emph{{shiny: Web Application Framework for R}}. \url{https://CRAN.R-project.org/package=shiny}.

\leavevmode\vadjust pre{\hypertarget{ref-Chatfield1985TIEo}{}}%
Chatfield, C. 1985. {``The Initial Examination of Data.''} \emph{Journal of the Royal Statistical Society. Series A. General} 148 (3): 214--53.

\leavevmode\vadjust pre{\hypertarget{ref-eliznlsy}{}}%
Cooksey, Elizabeth C. 2017. {``Using the National Longitudinal Surveys of Youth ({NLSY}) to Conduct Life Course Analyses.''} In \emph{Handbook of Life Course Health Development}, edited by Richard M. Lerner Neal Halfon Christoper B. Forrest, 561--77. Cham: Springer. https://doi.org/\url{https://doi.org/10.1007/978-3-319-47143-3_23}.

\leavevmode\vadjust pre{\hypertarget{ref-DasuTamraparni2003Edma}{}}%
Dasu, Tamraparni, and Theodore Johnson. 2003. \emph{Exploratory Data Mining and Data Cleaning}. Wiley Series in Probability and Statistics. Hoboken: WILEY.

\leavevmode\vadjust pre{\hypertarget{ref-janitor}{}}%
Firke, Sam. 2020. \emph{{janitor: Simple Tools for Examining and Cleaning Dirty Data}}. \url{https://CRAN.R-project.org/package=janitor}.

\leavevmode\vadjust pre{\hypertarget{ref-racismnotrace}{}}%
Fullilove, M. T. 1998. {``Comment: Abandoning ``Race" as a Variable in Public Health Research--an Idea Whose Time Has Come.''} \emph{American Journal of Public Health} 88 (9): 1297--98.

\leavevmode\vadjust pre{\hypertarget{ref-grimshaw}{}}%
Grimshaw, Scott D. 2015. {``A Framework for Infusing Authentic Data Experiences Within Statistics Courses.''} \emph{The American Statistician} 69 (4): 307--14. https://doi.org/\url{https://doi.org/10.1080/00031305.2015.1081106}.

\leavevmode\vadjust pre{\hypertarget{ref-SAGER}{}}%
Heidari, S., T. F. Babor, P. De Castro, S. Tort, and M. Curno. 2016. {``Sex and Gender Equity in Research: Rationale for the {SAGER} Guidelines and Recommended Use.''} \emph{Research Integrity and Peer Review} 1 (2). https://doi.org/\url{https://doi.org/10.1186/s41073-016-0007-6}.

\leavevmode\vadjust pre{\hypertarget{ref-purrr}{}}%
Henry, Lionel, and Hadley Wickham. 2020. \emph{{purrr: Functional Programming Tools}}. \url{https://CRAN.R-project.org/package=purrr}.

\leavevmode\vadjust pre{\hypertarget{ref-penguins-data}{}}%
Horst, Allison Marie, Alison Presmanes Hill, and Kristen B Gorman. 2020. \emph{Palmerpenguins: Palmer Archipelago (Antarctica) Penguin Data}. \url{https://doi.org/10.5281/zenodo.3960218}.

\leavevmode\vadjust pre{\hypertarget{ref-HuebnerMariannePhD2016Asat}{}}%
Huebner, Marianne, Werner Vach, and Saskia le Cessie. 2016. {``A Systematic Approach to Initial Data Analysis Is Good Research Practice.''} \emph{The Journal of Thoracic and Cardiovascular Surgery} 151 (1): 25--27.

\leavevmode\vadjust pre{\hypertarget{ref-ilk2004}{}}%
Ilk, Ozlem. 2004. {``Exploratory Multivariate Longitudinal Data Analysis and Models for Multivariate Longitudinal Binary Data.''} PhD thesis, Iowa State University. \url{https://doi.org/10.31274/rtd-180813-11012}.

\leavevmode\vadjust pre{\hypertarget{ref-kennedy2020using}{}}%
Kennedy, Lauren, Katharine Khanna, Daniel Simpson, and Andrew Gelman. 2020. {``Using Sex and Gender in Survey Adjustment.''} \url{https://arxiv.org/abs/2009.14401}.

\leavevmode\vadjust pre{\hypertarget{ref-tamedata}{}}%
Kim, A. Y, C. Ismay, and J. Chunn. 2018. {``The Fivethirtyeight {R} Package: ``Tame Data" Principles for Introductory Statistics and Data Science Courses.''} \emph{Technology Innovations in Statistics Education} 11 (1). https://doi.org/\url{https://doi.org/10.5070/T511103589}.

\leavevmode\vadjust pre{\hypertarget{ref-KollerManuel2016rARP}{}}%
Koller, Manuel. 2016. {``{robustlmm: An R Package for Robust Estimation of Linear Mixed-Effects Models}.''} \emph{Journal of Statistical Software} 75 (6): 1--24.

\leavevmode\vadjust pre{\hypertarget{ref-notaverage}{}}%
Moncrief, Marc. 2015. {``By the Numbers - the Average Australian Doesn't Exist ... Not a Single One of Us Is 'Normal'.''} \url{https://bit.ly/smh-not-normal}.

\leavevmode\vadjust pre{\hypertarget{ref-doh}{}}%
Office of Management and Budget. 1997. {``Revisions to the Standards for the Classification of Federal Data on Race and Ethnicity.''} \url{https://www.govinfo.gov/content/pkg/FR-1997-10-30/pdf/97-28653.pdf}.

\leavevmode\vadjust pre{\hypertarget{ref-opendata}{}}%
Open Knowledge Foundation. 2021. {``Open Definition. Defining Open in Open Data, Open Content, and Open Knowledge.''} 2021. \url{http://opendefinition.org/od/2.1/en/}.

\leavevmode\vadjust pre{\hypertarget{ref-patchwork}{}}%
Pedersen, Thomas Lin. 2020. \emph{{patchwork: The Composer of Plots}}. \url{https://CRAN.R-project.org/package=patchwork}.

\leavevmode\vadjust pre{\hypertarget{ref-MichaelRPergamit2001DWTN}{}}%
Pergamit, Michael R., Charles R. Pierret, Donna S. Rothstein, and Jonathan R. Veum. 2001. {``Data Watch: The National Longitudinal Surveys.''} \emph{The Journal of Economic Perspectives} 15 (2): 239--53.

\leavevmode\vadjust pre{\hypertarget{ref-R}{}}%
R Core Team. 2020. \emph{R: A Language and Environment for Statistical Computing}. Vienna, Austria: R Foundation for Statistical Computing. \url{https://www.R-project.org/}.

\leavevmode\vadjust pre{\hypertarget{ref-SingerJudithD2003Alda}{}}%
Singer, Judith D, and John B Willett. 2003. \emph{Applied Longitudinal Data Analysis: Modeling Change and Event Occurrence}. Oxford u.a: Oxford Univ. Pr.

\leavevmode\vadjust pre{\hypertarget{ref-notiris}{}}%
Stodel, Megan. 2020. {``Stop Using Iris.''} \url{https://www.meganstodel.com/posts/no-to-iris/}.

\leavevmode\vadjust pre{\hypertarget{ref-brolgar}{}}%
Tierney, Nicholas, Di Cook, and Tania Prvan. 2020. \emph{{brolgar: BRowse Over Longitudinal data Graphically and Analytically in R}}. \url{https://github.com/njtierney/brolgar}.

\leavevmode\vadjust pre{\hypertarget{ref-tukey}{}}%
Tukey, John W. (John Wilder). 1977. \emph{Exploratory Data Analysis}. Addison-Wesley Series in Behavioral Science. Reading, Mass.: Addison-Wesley Pub. Co.

\leavevmode\vadjust pre{\hypertarget{ref-validate}{}}%
van der Loo, Mark P. J., and Edwin de Jonge. 2021. {``Data Validation Infrastructure for {R}.''} \emph{Journal of Statistical Software} 97 (10): 1--31. \url{https://doi.org/10.18637/jss.v097.i10}.

\leavevmode\vadjust pre{\hypertarget{ref-LooMarkvander2018Sdcw}{}}%
van der Loo, Mark, and Edwin de Jonge. 2018. \emph{Statistical Data Cleaning with Applications in r}.

\leavevmode\vadjust pre{\hypertarget{ref-mass}{}}%
Venables, W. N., and B. D. Ripley. 2002. \emph{Modern Applied Statistics with s}. Fourth. New York: Springer. \url{http://www.stats.ox.ac.uk/pub/MASS4}.

\leavevmode\vadjust pre{\hypertarget{ref-tsibble}{}}%
Wang, Earo, Dianne Cook, and Rob J Hyndman. 2020. {``A New Tidy Data Structure to Support Exploration and Modeling of Temporal Data.''} \emph{Journal of Computational and Graphical Statistics} 29 (3): 466--78. \url{https://doi.org/10.1080/10618600.2019.1695624}.

\leavevmode\vadjust pre{\hypertarget{ref-plyr}{}}%
Wickham, Hadley. 2011. {``The Split-Apply-Combine Strategy for Data Analysis.''} \emph{Journal of Statistical Software, Articles} 40 (1): 1--29. https://doi.org/\url{https://doi.org/10.18637/jss.v040.i01}.

\leavevmode\vadjust pre{\hypertarget{ref-WickhamHadley2014TD}{}}%
---------. 2014. {``Tidy Data.''} \emph{Journal of Statistical Software} 59 (10): 1--23.

\leavevmode\vadjust pre{\hypertarget{ref-ggplot2}{}}%
---------. 2016. \emph{{ggplot2: Elegant Graphics for Data Analysis}}. Springer-Verlag New York. \url{https://ggplot2.tidyverse.org}.

\leavevmode\vadjust pre{\hypertarget{ref-stringr}{}}%
---------. 2019. \emph{{stringr: Simple, Consistent Wrappers for Common String Operations}}. \url{https://CRAN.R-project.org/package=stringr}.

\leavevmode\vadjust pre{\hypertarget{ref-tidyr}{}}%
---------. 2020. \emph{{tidyr: Tidy Messy Data}}. \url{https://CRAN.R-project.org/package=tidyr}.

\leavevmode\vadjust pre{\hypertarget{ref-tidyverse}{}}%
Wickham, Hadley, Mara Averick, Jennifer Bryan, Winston Chang, Lucy D'Agostino McGowan, Romain François, Garrett Grolemund, et al. 2019. {``Welcome to the {tidyverse}.''} \emph{Journal of Open Source Software} 4 (43): 1686. \url{https://doi.org/10.21105/joss.01686}.

\leavevmode\vadjust pre{\hypertarget{ref-dplyr}{}}%
Wickham, Hadley, Romain François, Lionel Henry, and Kirill Müller. 2020. \emph{{dplyr: A Grammar of Data Manipulation}}. \url{https://CRAN.R-project.org/package=dplyr}.

\leavevmode\vadjust pre{\hypertarget{ref-readr}{}}%
Wickham, Hadley, and Jim Hester. 2020. \emph{{readr: Read Rectangular Text Data}}. \url{https://CRAN.R-project.org/package=readr}.

\leavevmode\vadjust pre{\hypertarget{ref-nlsy79edu}{}}%
Wolpin, Kenneth I. 2005. {``National Longitudinal Survey of Youth 1979 Cohort, 1979-2016 (Rounds 1-28).''} Published by Bureau of Labor Statistics, U.S. Department of Labor. \url{https://www.bls.gov/opub/mlr/2005/02/art3full.pdf}.

\leavevmode\vadjust pre{\hypertarget{ref-knitr}{}}%
Xie, Yihui. 2014. {``Knitr: A Comprehensive Tool for Reproducible Research in {R}.''} In \emph{Implementing Reproducible Computational Research}, edited by Victoria Stodden, Friedrich Leisch, and Roger D. Peng. Chapman; Hall/CRC. \url{http://www.crcpress.com/product/isbn/9781466561595}.

\leavevmode\vadjust pre{\hypertarget{ref-rmarkdown}{}}%
Xie, Yihui, Christophe Dervieux, and Emily Riederer. 2020. \emph{R Markdown Cookbook}. Boca Raton, Florida: Chapman; Hall/CRC. \url{https://bookdown.org/yihui/rmarkdown-cookbook}.

\leavevmode\vadjust pre{\hypertarget{ref-kableExtra}{}}%
Zhu, Hao. 2019. \emph{{kableExtra: Construct Complex Table with 'kable' and Pipe Syntax}}. \url{https://CRAN.R-project.org/package=kableExtra}.

\end{CSLReferences}

\bibliographystyle{unsrt}
\bibliography{references.bib}

\end{document}